\documentclass[a4paper,11pt,twoside]{article}
\usepackage[T2A]{fontenc}
\usepackage{mpcs}
\usepackage{rumathbr}
\usepackage{hyperref}

\renewcommand{\ep}{1}
\begin{document}

	\vestnik{2024}{4}
	
	\setcounter{doi}{4}
	
	\setcounter{page}{39}

%%\begin{document}

%%\renewcommand{\ep}{1}
\selectlanguage{english}

\udk{524.7-8}
\bbk{32.973-018}
\art[Efficiency of parallel computations of gravitational forces]%
{Efficiency of parallel computations of gravitational forces by TreeCode method in N-body models}%
{Эффективность параллельных вычислений гравитационных сил методом TreeCode в моделях N-тел}
%\support{Grant support information}
 \support{This work supported by the Russian Science Foundation~(grant no. 23-71-00016, \\  https://rscf.ru/project/23-71-00016/).}

\fio[Николай Михайлович Кузьмин]{Kuzmin}{Nikolay}{M.}
%\sdeg{Candidate of Physical and Mathematical Sciences}{кандидат физико-математических наук}
\pos{Department of Information Systems and Computer Simulations}{доцент кафедры информационных систем и компьютерного моделирования}
\emp{Volgograd State University}{Волгоградский государственный университет}
\email{nikolay.kuzmin@volsu.ru}
\orcid{0000-0003-4074-0970}
\addr{Prosp. Universitetsky, 100, 400062 Volgograd, Russian Federation}{просп. Университетский, 100, 400062 г. Волгоград, Российская Федерация}

\fio[Данила Сергеевич Сиротин]{Sirotin}{Danila}{S.}
%\sdeg{Candidate of Physical and Mathematical Sciences}{Candidate of Physical and Mathematical Sciences}
\pos{Department of Information Systems and Computer Simulations}{ассистент кафедры информационных систем и компьютерного моделирования}
\emp{Volgograd State University}{Волгоградский государственный университет}
\email{d.sirotin@volsu.ru}
\orcid{0000-0001-8956-570X}
\addr{Prosp. Universitetsky, 100, 400062 Volgograd, Russian Federation}{просп. Университетский, 100, 400062 г. Волгоград, Российская Федерация}

\fio[Александр Валентинович Хоперсков]{Khoperskov}{Alexander}{V.}
%\sdeg{Doctor of Physical and Mathematical Sciences}{доктор физико-математических наук}
\pos{Department of Information Systems and Computer Simulations}{профессор кафедры информационных систем и компьютерного моделирования}
\emp{Volgograd State University}{Волгоградский государственный университет}
\email{khoperskov@volsu.ru}
\orcid{0000-0003-0149-7947}
\addr{Prosp. Universitetsky, 100, 400062 Volgograd, Russian Federation}{просп. Университетский, 100, 400062 г. Волгоград, Российская Федерация}

\maketitle

\noindent \underline{\bf Citation}: Kuzmin N.M., Sirotin D.S., Khoperskov A.V. Efficiency of parallel computations of gravitational forces by TreeCode method in N-body models // Mathematical Physics and Computer Simulation, 2024, vol. 27, no. 4, pp. 39-55. \\ 
\href{https://mp.jvolsu.com/index.php/en/archive-en}{https://mp.jvolsu.com/index.php/en/archive-en} \\ 
https://doi.org/10.15688/mpcm.jvolsu.2024.4.4  

\begin{abstract}
Modeling of collisionless galactic systems is based on the N-body model, which requires large computational resources due to the long-range nature of gravitational forces. The most common method for calculating gravity is the TreeCode algorithm, which provides a faster calculation of the force compared to the direct summation of contributions from all particles for N-body simulation. An analysis of the computational efficiency is performed for models with the number of particles up to $10^{8}$. We considered several processors with different architectures in order to determine the performance of parallel simulations based on the OpenMP standard. An analysis of the use of extra threads in addition to physical cores shows an increase in simulation performance only when all logical threads are loaded, which doubles the total number of threads. This gives an increase in the efficiency of parallel computing by 20 percent on average.
\end{abstract}

{параллельные вычисления, гравитирующие системы, OpenMP, архитектура процессоров, технология Hyper-Threading}

\section{Introduction}

The N-body model is used to simulate the dynamics of galactic systems, including open clusters and globular
clusters~\cite{Ishchenko-etal-2024globular-clusters, Bissekenov-etal-2024open-cluster}, single galaxies, interacting galaxies,
groups and clusters of galaxies~\cite{Athanassoula-1998N-Body, Kulikov-2016interaction-galaxies,
Ciambur-etal-2021N-body-bar, Titov-Khoperskov-2022Peter, Bagla-2005Cosmological-N-Body, Nipoti-2021N-body, Khoperskov-etal-2024Galaxies-cE,
Tikhonenko-etal-2021Sotnikova, Just-etal-2023Polyachenko, Fridman-Khoperskov-2012book}. There are a significant number of numerical
algorithms for parallel implementation of methods for calculating gravitational forces in N-body system. The simplest direct summation
method of each particle to each other one (Particle-Particle, PP) greatly limits the number of particles due to the complexity of $O(N^2)$,
although it can serve as a tool for testing~\cite{Fridman-Khoperskov-2012book}. The algorithm proposed in~\cite{Barnes-Hut-1986treecode}
based on a special partitioning of the volume with particles is traditionally called TreeCode. It reduces the complexity of calculations
to ${O(N\,\log(N))}$ by reducing the accuracy of the contribution to gravity from distant particles.

A variety of fast methods for calculating the gravitational force in a system of $N$ particles have been proposed over the past decades. We highlight such methods as Fast Fourier Transform (FFT) with complexity close to ${O(N\,\log(N))}$, Particle-Mesh Method, Particle-Particle + Particle-Mesh (${\rm P^3M}$) and others~\cite{Walther-2003P3M, Kyziropoulos-etal-2015Parallel-N-Body, Bagla-2005Cosmological-N-Body}. The FFT method has a number of advantages for simulating modified gravity~\cite{Ruan-etal-2022N-body-modified-gravity}. A special feature of the Grid-of-Oct-Trees-Particle-Mesh (GOTPM) algorithm is a hybrid scheme combining particle-mesh (PM) and oct-trees of Barnes-Hut method~\cite{Dubinski-etal-2004treecode, Huillier-etal-2014N-body}. A generalization of tree codes is Fast Multipole Method with number of computational operations $O(N)$~\cite{Yokota-Barba-2011Fast-Multipole-Method, Bagla-2005Cosmological-N-Body}.

%\cite{Liu-Bhatt-2000parallel-N-body}

Large cosmological simulations projects aim to study the cosmological evolution of dark and baryonic matter, as well as the physics of galaxy clusters over $10^{10}$ years based on N-body and/or hydrodynamical simulations~\cite{Potter-etal-2017PKDGRAV3, Vogelsberger-etal-2020Cosmological-simulations}. Examples of such projects are MillenniumTNG (3000-megaparsec box with more than 1 trillion particles)~\cite{Hernandez-Aguayo-etal-2023MillenniumTNG}, PKDGRAV3 (${N = 8 \cdot 10^{12}}$)~\cite{Potter-etal-2017PKDGRAV3}, EAGLE simulation (Evolution and Assembly of GaLaxies and their Environments)~\cite{Schaye-etal-2015EAGLE-project}, FLAMINGO project~\cite{Schaye-etal-2023cosmolog}, Illustris-TNG~\cite{Pillepich-etal-2018Illustris-TNG} and many others.

%SWIFT

The use of cosmological models has proven its effectiveness in solving problems of the dynamics of single galaxies or galaxy groups, as in the Auriga project~\cite{Grand-etal-2016AURIGA-project} and the Apostle project~\cite{Fattahi-etal-2016Apostle-project}. The use of high-order parallel N-body codes for GPUs allows one to study the dynamics of globular clusters~\cite{Ishchenko-etal-2024Berczik} and compact elliptical galaxies on cosmological time scale~\cite{Khoperskov-etal-2024Galaxies-cE}. 

The number of real stars in galaxies is always much larger than the number of particles in the model, and this requires the fulfillment of the collisionless condition of the stellar component in the simulations~\cite{Smirnov-etal-2017slow-bars, Khrapov-etal-2023dwarf}.
Various approaches have been proposed to improve the quality of N-body models. An example is the GalaxyFlow, which aims to create a method to go from very coarse stellar phase-space density in the original numerical simulations to more accurate stellar phase-space density estimates for astronomical applications~\cite{Lim-etal-2024GalaxyFlow}.

The aim of the work is to calculate the efficiency of parallel N-body simulations based on TreeCode. Since the bottleneck of the Message Passing Interface is the speed of data exchange between processors, our analysis is limited to the OpenMP standard within the shared memory.
Processors have both physical cores (processor cores) and extra threads (virtual or logical cores) thanks to Hyper-Threading Technology.
We present the details of our analysis of the performance of Treecode for calculating gravitational forces using extra threads in addition to the processor cores.

\section{N-body simulation with TreeCode} 

The dynamics of $N$ gravitating particles is described by a system of equations
\begin{equation}
\left\{
\begin{aligned}
\frac{d\mathbf{r}_i}{dt} &=  \mathbf{v}_i,\\
\frac{d\mathbf{v}_i}{dt} &= \sum_{j \neq i}^N G\,m_j\,\frac{\mathbf{r}_j - \mathbf{r}_i}{\left(\left|\mathbf{r}_j - \mathbf{r}_i\right|^2 + \varepsilon^2\right)^{3/2}},
\end{aligned}
\right.
\label{eq:equations-motion-N-body}
\end{equation}
where $\mathbf{r}_i$, $\mathbf{v}_i$ and $m_i$ are the radius vector, velocity and mass of the $i$-th particle, respectively,
%${G \simeq 6.6743 \times 10^{-11}~\text{м}^3/(\text{кг}\cdot \text{с}^2)}$~--- гравитационная постоянная.
$G$ is the gravitational constant, $\varepsilon$ is the smoothing radius that gives a collisionless system.
The dynamical N-body model includes only a stellar disc and lacks a dark halo defined by a given gravitational potential, as in~\cite{Butenko-etal-2022MPCS}.
The initial equilibrium in the rotating disc is provided by the balance of gravitational and centrifugal forces with an additional contribution from random particle motion, which is described by spatial distributions of velocity dispersions in three directions~\cite{Fridman-Khoperskov-2012book, Pejch-etal-2023, Khoperskov-etal-2024Galaxies-cE, Khrapov-Khoperskov-2024Frontiers}. The dispersion of the random velocity component along the vertical coordinate is an analogue of pressure and compensates for the gravitational force, which gives an equilibrium disk in the transverse direction.   

%Для численных расчетов систему~\eqref{eq:equations-motion-N-body} обычно переписывают в виде
%\begin{equation}
%\left\{
%\begin{aligned}
%\frac{d\mathbf{r}_i}{dt} &=  \mathbf{v}_i,\\
%\frac{d\mathbf{v}_i}{dt} &= \sum_{j \neq i}^N G\,m_j\,\frac{\mathbf{r}_j - \mathbf{r}_i}{\left(\left|\mathbf{r}_j - \mathbf{r}_i\right|^2 + \varepsilon^2\right)^{3/2}},
%\end{aligned}
%\right.
%\label{eq:equations-motion-N-body-smooth}
%\end{equation}

The numerical integration of the system~\eqref{eq:equations-motion-N-body} is based on the classical second-order leap-frog scheme:
\begin{equation}
\mathbf{v}^{n+1/2}_i = \mathbf{v}^n_i + \frac{\tau}{2}\, \mathbf{a}_i^n,
\label{eq:numerical-v-pred}
\end{equation}
\begin{equation}
\mathbf{r}^{n+1}_i = \mathbf{r}^n + \tau\, \mathbf{v}^{n+1/2}_i,
\label{eq:numerical-r}
\end{equation}
\begin{equation}
%\mathbf{a}^{n+1}_i = \sum_{j \neq i}^N G\,m_j\,\frac{\mathbf{r}^{n+1}_j - \mathbf{r}^{n+1}_i}{\left|\mathbf{r}^{n+1}_j - \mathbf{r}^{n+1}_i\right|^3},
\mathbf{a}^{n+1}_i = \mathbf{a}(\mathbf{r}^{n+1}),
\label{eq:numerical-accelerations}
\end{equation}
\begin{equation}
\mathbf{v}^{n+1}_i = \mathbf{v}^{n+1/2}_i + \frac{\tau}{2}\, \mathbf{a}_i^{n+1},
\label{eq:numerical-v-corr}
\end{equation}
where $\tau$ is the time step, $\mathbf{a}_i^n$ is the acceleration of the $i$-th particle, index $n$ corresponds to the time $t_n$.

Direct summation in the formula~\eqref{eq:equations-motion-N-body} for each particle of the contributions from all other bodies (Particle-Particle, PP method) yields a computational complexity of $O(N^2)$. The quadratic dependence of the computational time on the number of particles greatly limits such a naive approach. However, the direct method has a number of advantages. Obviously, the calculation error for PP is minimal compared to approximate algorithms for the same $N$, which provides a significant advantage for satisfying the physical conservation laws~\cite{Khrapov-Khoperskov-2024Frontiers, Khrapov-etal-2018N-body}. This allows for a higher-quality reproduction of the evolution of a gravitating system over large times and small scales, which is especially important for the gas component with large mass density gradients~\cite{Khoperskov-etal-2024Galaxies-cE}.

The calculation of gravitational forces in (\ref{eq:equations-motion-N-body}) is performed by the TreeCode method~\cite{Barnes-Hut-1986treecode}, based on the idea of an approximate construction of the gravitational potential of  ``distant'' particles group. The contribution of such a group is calculated as for one particle with the total mass of the entire group, located at the center of mass. The contribution of ``near'' particles is calculated by an exact formula. The implementation of this approach is based on a hierarchical partitioning of the space into eight octants until there is no more than one particle left in each subdomain. The division into ``near'' and ``distant'' groups is determined by the value of the parameter $\theta$. It has the meaning of the solid angle at which a given subdomain is visible from the point under consideration. This method has an algorithmic complexity ${O(N\, \log(N))}$. The TreeCode method turns into a direct method PP for calculating accelerations
 with algorithmic complexity $O(N^2)$ for ${\theta \to 0}$.

The key problem is the choice of the criterion when the group of particles is distant~\cite{Springel-etal-2001GADGET, Bagla-2005Cosmological-N-Body}. Since the tree code characteristics, the errors in calculating the gravitational forces and the features of the spatial distribution of particles are interconnected.

Calculating gravitational forces takes up the lion's share of the processor time regardless of the method of numerical integration of the equations system. Using the TreeCode method in this work requires over 95 percent of the time to calculate accelerations. This value varies within $2-3$ percent depending on the choice of the computing system.

%=======================================================
\section{Code performance analysis}

We use the FORTRAN-77 version of TreeCode\footnote{https://home.ifa.hawaii.edu/users/barnes/treecode_old/index.html} adapted for galactic stellar disc dynamics simulations~\cite{Fridman-Khoperskov-2012book}. This code is parallelized using OpenMP technology.

Using the compiler \mbox{\texttt{gfortran 12.2.0}} with the optimization switch \texttt{-O3} provides the most complete optimization in terms of execution time without specific extensions of the instruction set architecture (ISA) of the target processor. Aggressive optimization methods are not used, since they may affect the numerical results, such as option \texttt{-ffast-math}, which allows deviations from strict compliance with the  floating-point number representation standard IEEE~754, for example, in the case of subnormal numbers and the implementation of associativity of computations. 
This choice is due to the fact that we focused on the study of quantitative characteristics associated with parallel computing and do not consider a specific processor instruction set architecture. Solving mathematical simulation problems makes it preferable to use specific extensions of the instruction set of the target processor (the SSE family of extensions of various versions, AVX, etc.) using switch \texttt{-march=native}, which allows using all available instructions supported by the processor and the compiler.

We calculate the following characteristics of software parallelization performance: dependencies of the execution time of one step of the computational cycle $T$ (Time), the computations speedup $S$ (Speedup) and the efficiency of parallelization $E$ (Efficiency) on the number of threads used $M$. The definitions of these quantities are
\begin{equation}
S(M) = \frac{T(1)}{T(M)},
\quad
E(M) = \frac{S(M)}{M},
\label{eq:speed-eff}
\end{equation}
where $E$ is the efficiency without extra threads. 

The parameters $T$, $S$, and $E$ are measured on several different shared-memory computers (symmetric multiprocessing, SMP), the parameters of which are shown in Table~\ref{tbl:sys}. It is important to note that the processors are capable of dynamically changing the clock frequency depending on the current computing load level (Intel Turbo Boost and AMD Precision Boost technologies). Additionally, there is the ability to execute two logical computing threads on one physical processor core (Hyper-Threading Technology). The total number of available computing threads ${M = M_P + M_E}$ consists of physical cores $M_P$ and additional logical computing threads (extra threads) $M_E$. We compare two types of CPUs, both with support for hyper-threading technology (${M_E = M_P}$, ${M = 2M_P}$) and without it (${M_E = 0}$, ${M = M_P}$). Thus, threads on a single CPU may be nonequivalent, which is almost always a problem for a parallel program. Formulas for $E(M)$ in the form~\eqref{eq:speed-eff} are traditionally used for ${M \le M_P}$. The efficiency in the case of ${M > M_P}$ (${M_E \le 1}$) should be written in the form
\begin{equation}\label{eq:efficiency-MP}
    E^{(P)}(M) = \frac{S(M)}{M_P} = \frac{T(1)}{T(M)\, M_P},
\end{equation}
since only physical cores must be considered, ${M_P = M - M_E}$. 

\begin{table}[!htb] 
\caption{Characteristics of computing systems}
\label{tbl:sys}
{\small
\begin{center}
\begin{tabular}{|p{1.25cm}|p{4.8cm}|p{2.3cm}|p{2.5cm}|p{2.8cm}|}
\hline
Name & CPU & Total number of cores\newline (number of logical threads) & RAM & CPU release date\\
\hline
Beta & 2 $\times$ Intel Xeon E5405,\newline 2~GHz & 8 (8) & 16~GB\newline DDR2-667 & 2007\\
\hline
Epsilon & 2 $\times$ Intel Xeon E5540,\newline 2.53~GHz & 8 (16) & 16~GB\newline DDR3-1600 & 2009\\
\hline
K80 & 2 $\times$ Intel Xeon \mbox{E5-2687W v3},\newline 3.1~GHz (3.5~GHz with Turbo Boost) & 20 (40) & 512~GB\newline DDR3-2133 & 2014\\
\hline
M105 & Intel Core i5-6400,\newline 2.7~GHz (3.3~GHz with Turbo Boost) & 4 (4) & 8~GB\newline DDR3-1600 & 2015\\
\hline
M111 & Intel Core i5-9400,\newline 2.9~GHz (4.1~GHz with Turbo Boost) & 6 (6) & 16~GB\newline DDR4-2666 & 2019\\
\hline
M113 & 12th Gen Intel Core i5-12400F\newline 2.5~GHz (4.4~GHz with Turbo Boost) & 6 (12) & 16~GB\newline DDR5-5200 & 2022\\
\hline
M105S & Intel Core i9-9900KF\newline 3.6~GHz (5.0~GHz with Turbo Boost) & 8 (16) & 64~GB\newline DDR4-2400 & 2019\\
\hline
M301 & 2 $\times$ Intel Xeon E5-2630 v3,\newline 2.4~GHz (3.2~GHz with Turbo Boost) &  16 (32) & 16~GB\newline DDR4-1866 & 2014 \\ 
\hline
Ryzen & AMD Ryzen 7 2700,\newline 3.2~GHz (4.1~GHz with Precision Boost) &  8 (16) & 16~GB\newline DDR4-3200 & 2018 \\ 
\hline
\end{tabular}  
\end{center}
}
\end{table}

All results were obtained for averaged values of execution time of one step of the computational cycle $T$. Averaging was performed over ten steps of time integration. Therefore, we plot absolute errors in the figures as vertical bars. Measurement errors were calculated by using the following expression of standard error of the mean:
\begin{equation}
\Delta(T) = \sqrt{\frac{\sum_{i=1}^n\left(T_i - \langle T \rangle\right)^2}{n(n - 1)}},
\end{equation}
where $\langle T \rangle$ is the mean value of $T$, ${n = 10}$ is the number of measurements.
The corresponding errors of $S$ and $E$ have larger values because they are calculated as ratios of $A/B$, which gives
\begin{equation}
\Delta\left(\frac{A}{B}\right) = \frac{A \cdot \Delta(B) + B \cdot \Delta(A)}{B^2}.
\end{equation}

The main parameter determining the accuracy of the gravitational force calculation in the TreeCode algorithm is the opening angle $\theta$. This free parameter actually specifies the hierarchical structure of the grid on which the gravitational potential is calculated. All numerical experiments in this work are carried out for ${\theta = 1}$. This choice is typical in the practice of N-body simulations. Reducing $\theta$ greatly increases the time $T$, all other things being equal. There is a transition from TreeCode to PP in the limit ${\theta \to 0}$.

\begin{figure}[!htb]
\centering
\includegraphics[width = 0.95\textwidth]{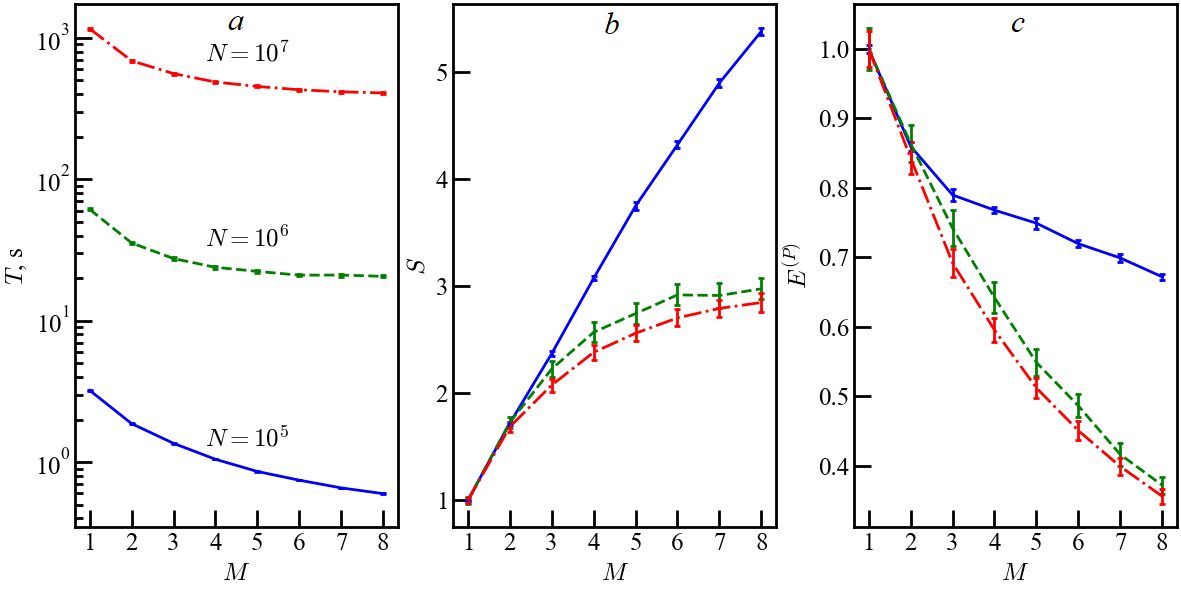}
\caption{
Computational time ($a$), speedup  ($b$) and efficiency  ($c$) for computer Beta in simulations with different numbers of particles: ${N = 10^5}$ (blue), ${N = 10^6}$ (green), ${N = 10^7}$ (red). 
 }
\label{fig:beta-M}
\end{figure}

\begin{figure}[!htb]
\centering
\includegraphics[width = 0.85\textwidth]{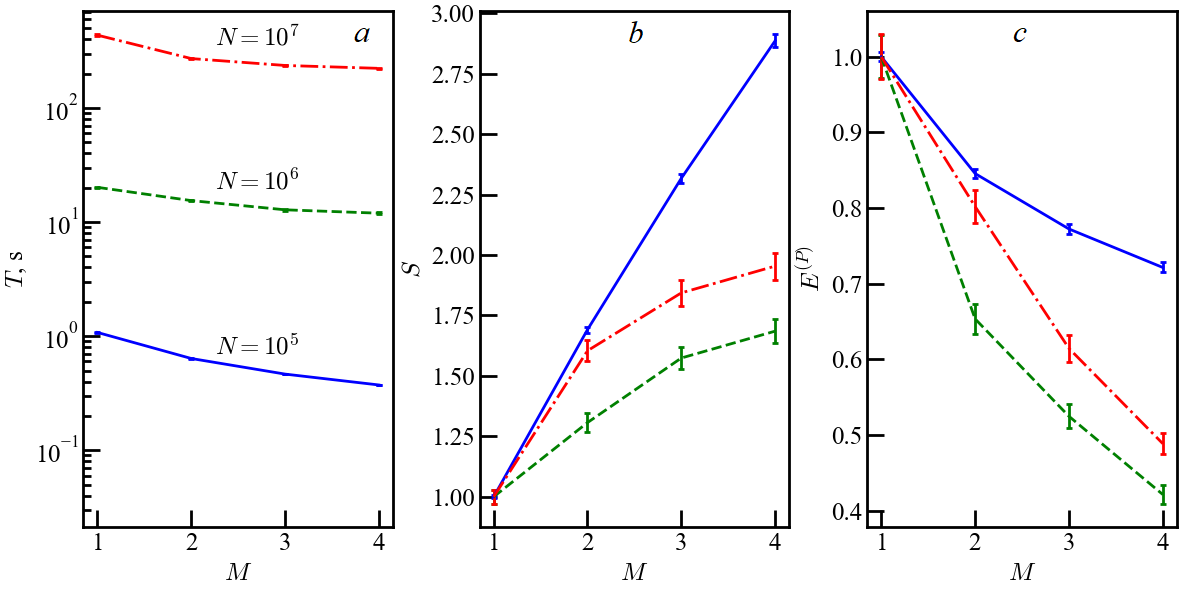}
\caption{
The same as in figure 1 for computer M105.  
  }
\label{fig:m105-M}
\end{figure}

\begin{figure}[!htb]
\centering
\includegraphics[width = 0.85\textwidth]{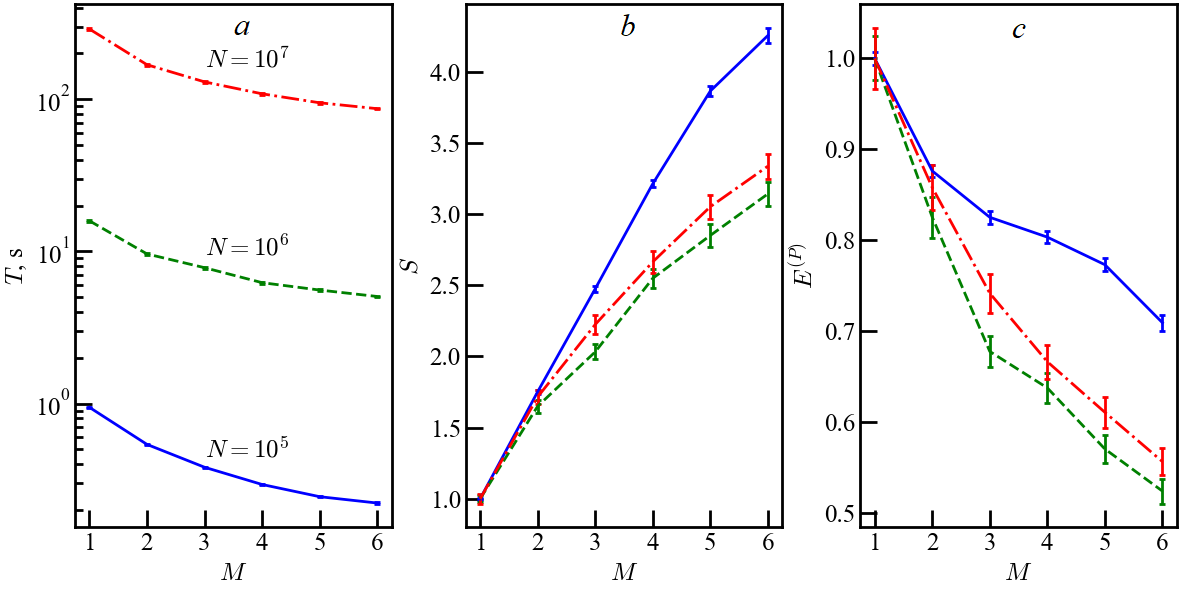}
\caption{
The same as in figure 1 for computer M111.  
  }
\label{fig:m111-M}
\end{figure}

\begin{figure}[!htb]
\centering
\includegraphics[width = 0.85\textwidth]{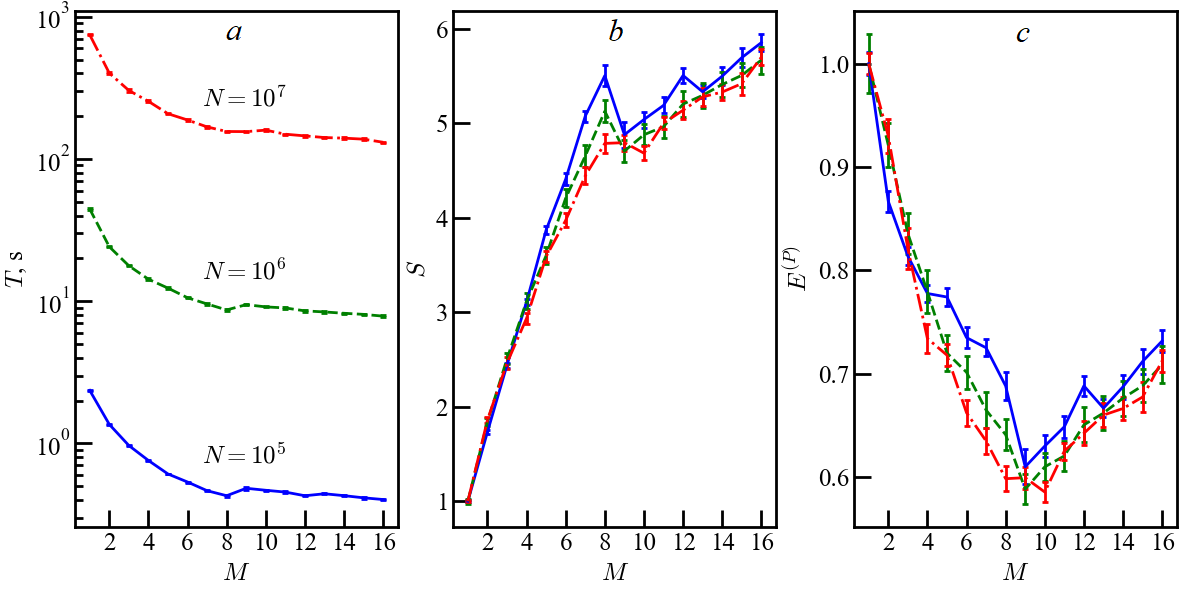}
\caption{
The same as in figure 1 for computer Epsilon.  
  }
\label{fig:epsilon-M}
\end{figure}

\begin{figure}[!htb]
\centering
\includegraphics[width = 0.95\textwidth]{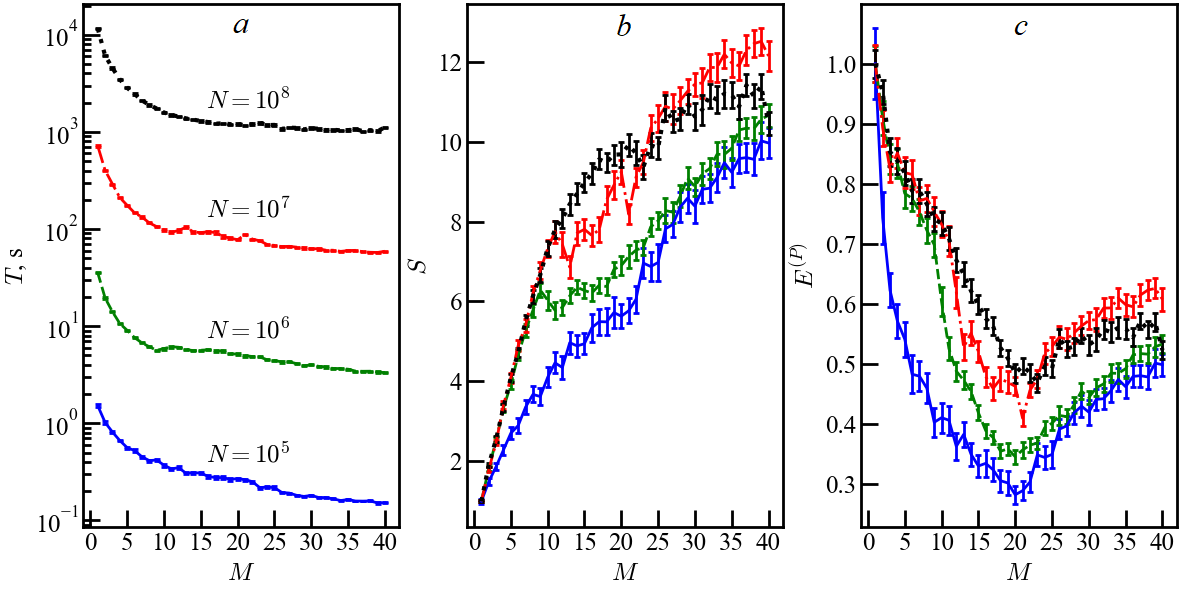}
\caption{
The same as in figure 1 for computer K80.  
  }
\label{fig:k80-M}
\end{figure}

\begin{figure}[!htb]
	\centering
	\includegraphics[width = 0.9\textwidth]{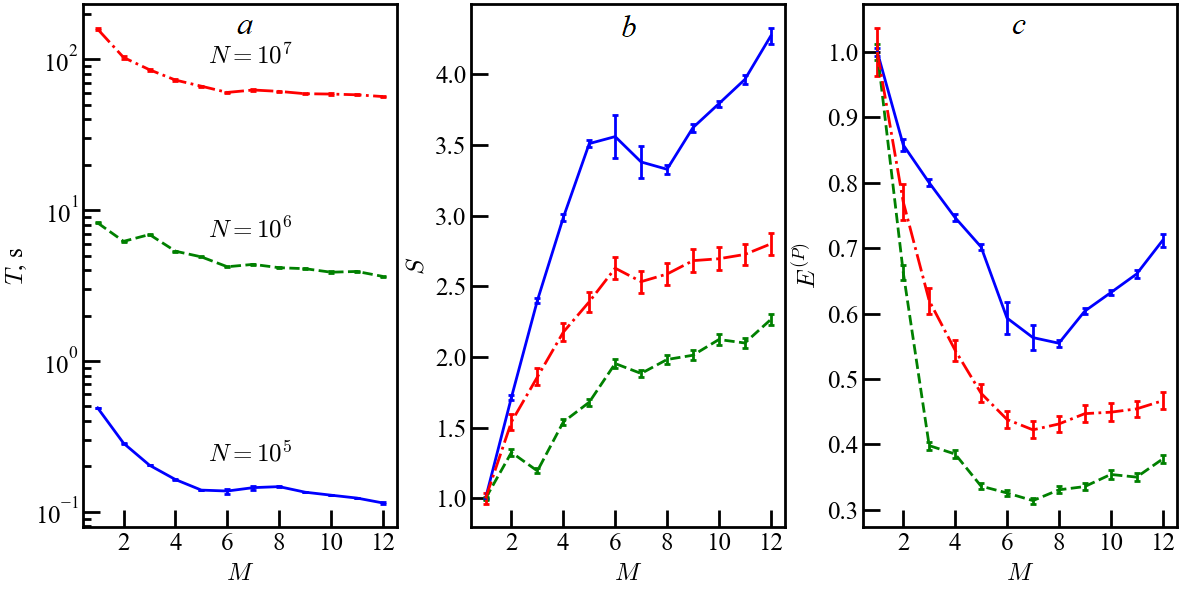}
	\caption{
		The same as in figure 1 for computer M113.
	}
	\label{fig:m113-M}
\end{figure}

\begin{figure}[!htb]
\centering
\includegraphics[width = 0.9\textwidth]{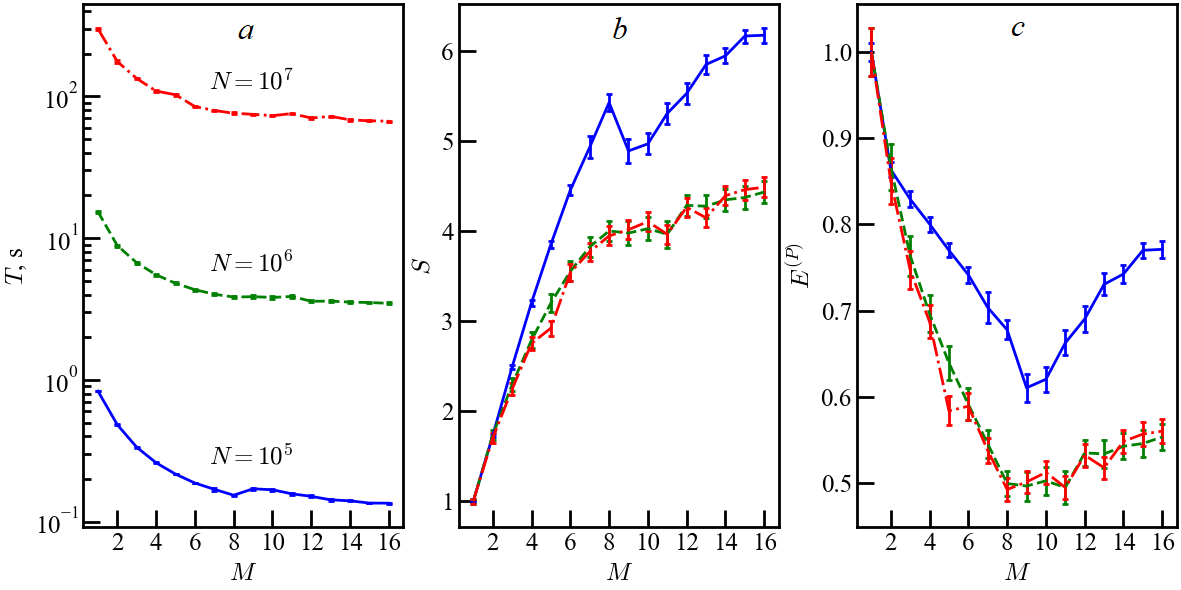}
\caption{
The same as in figure 1 for computer M105S. 
 }
\label{fig:m105s-M}
\end{figure}

\begin{figure}[!htb]
\centering
\includegraphics[width = 0.9\textwidth]{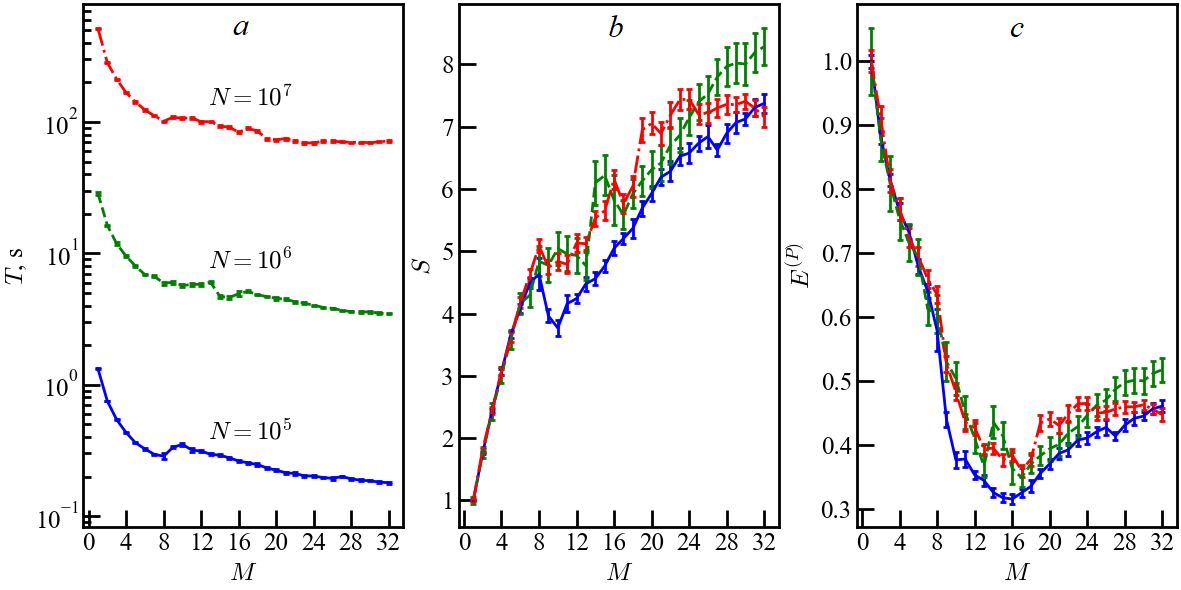}
\caption{
The same as in figure 1 for computer M301. 
 }
\label{fig:m301-M}
\end{figure}

The results of calculating the average time $T$, speedup $S$ and efficiency $E^{(P)}$ for different computing systems and the number of threads $M$ are shown in figures~\ref{fig:beta-M} --- \ref{fig:m301-M}. The simulation time $T$ consists of the execution time of the following components of the algorithm in accordance with formulas~\eqref{eq:numerical-v-pred} --- \eqref{eq:numerical-v-corr}: calculating gravitational forces for a system of $N$ particles ($T^{(grav)}$), determining velocities at the predictor stage ($T^{({\bf v}1)}$), new coordinates ($T^{({\bf r})}$), velocities at the corrector stage ($T^{({\bf v}2)}$) and accelerations ($T^{({\bf a})}$) of particles. Since ${T^{(grav)}/(T^{(grav)} + T^{({\bf v}1)} + T^{({\bf r})} + T^{({\bf v}2)} +T^{({\bf a})}) \simeq 0.96}$, we restrict ourselves to the approximation ${T = T^{(grav)}}$ below.  

Computer Beta has two processors with four physical cores each, which gives ${M_P = 8}$ and ${M_E = 0}$. The dependencies $T(M)$, $S(M)$, $E^{(P)}(M)$ for the models with ${N = 10^5}$, ${N = 10^6}$ and ${N = 10^7}$ are shown in figure~\ref{fig:beta-M}. The computation time $T$ decreases noticeably with the increase of the number of cores $M$ for ${N = 10^5}$. Experiments with ${N = 10^6}$ and ${N = 10^7}$ give a slower decrease with an exit to an almost constant level. This, respectively, affects the speedup $S(M)$ and the efficiency $E^{(P)}(M)$. The speedup reaches a plateau at ${M \to 8}$, in contrast to the simulation with ${N = 10^5}$ with an almost linear increase in speedup. The efficiency turns out to be worse than 0.4 when using all available physical cores in models with ${N \gtrsim 10^6}$.

Other processors without extra threads (${M_E = 0}$) have similar dependences of characteristics on the number of cores. The M105 processor is more modern and has a higher clock frequency than Beta. Therefore, the value of $T$ on computer M105 is approximately 3 times less than on Beta (figure~\ref{fig:m105-M}). However, the speedup and efficiency are better only within 10--15\%. Such results are also preserved for the more modern processor of M111 in figure~\ref{fig:m111-M}. Moreover, the growth of speedup continues to the maximum value of ${M_P = 6}$. There are higher speedups and efficiency for ${N \ge 10^7}$ in contrast to computer Beta due to the more modern architecture.

Next, we consider computing systems with ${M_E = M_P}$, which doubles the number of threads. Computer Epsilon has 8 additional logical threads and ${M = 16}$. The value of ${T(M = 1)}$ decreases compared to Beta due to the more advanced processor architecture (figure~\ref{fig:epsilon-M}). Using only physical cores (${M\le 8}$) is more efficient for Epsilon than in the case of Beta. The dependences of speedup and efficiency on the number of particles are significantly weakened (figure~\ref{fig:epsilon-M}). The difference in results for all three experiments with ${N = 10^5}$, ${N = 10^6}$ and ${N = 10^7}$ is small. All dependencies change sharply for the number of threads ${M > 8}$ when using extra threads. Two characteristic features stand out in parallel computing with ${M > M_P = 8}$.

First, adding several extra threads leads to a noticeable decrease in speedup and efficiency. For example, using extra threads with ${M_E = 1-2}$ increases the time $T$ compared to ${M = M_P = 8}$ (See figure \ref{fig:epsilon-M}$a$). Characteristic kinks in the dependencies $S(M)$ and $E^{(P)}(M)$ in the vicinity of ${M = M_P = 8}$ show the negative impact of a small number of extra threads on the overall efficiency of simulations. The second feature appears when using the maximum possible number of extra threads. Such a doubling of the number of threads to ${M = 16}$ allows exceeding the indicators with ${M = M_P = 8}$. Comparing the efficiencies of $E^{(P)}(8)$ and $E^{(P)}(16)$ yields an increase of 6 percent for ${N = 10^5}$ and 19 percent for ${N = 10^7}$ for computer Epsilon.

The nature of the dependencies $T(M)$, $S(M)$, $E^{(P)}(M)$ for ${M \le 8}$ for computer Epsilon (See figure~\ref{fig:epsilon-M}) is due to the fact that one logical computing thread is used on one physical processor core. A further increase in threads (${M > 8}$) leads to the use of two threads on one core, which violates the monotonic dependence. Computer Beta does not have extra threads (${M = M_P}$), which gives monotonicity of efficiency (see figure~\ref{fig:beta-M}).

Next, we consider computer K80 (See Table 1), including an additional model with ${N = 10^8}$, which requires more RAM (${\approx 6}$~GB). The total number of threads reaches ${M = 40}$ for ${M_P = 20}$ and ${M_E = 20}$. K80 contains 2 processors with shared memory. Figure \ref{fig:k80-M} shows our results for K80. There are peculiarities when the number of threads passes near ${M \simeq 10}$ and ${M = M_P = 20}$. The first is due to the two processors on K80. The transition from ${M = 10}$ to a higher number of threads means either using the second processor or/and partial calculations on extra threads of the first or second processor. The choice of a specific operating mode is determined by the operating system and is not adjustable by the user.

The internal scheduler does not provide information on the distribution of resources between devices. However, the dependencies $T(M)$, $S(M)$, $E^{(P)}(M)$ allow us to assume that the distribution of computing resources occurs as follows. At the beginning, one logical computing thread is allocated to one physical core of the first processor (${M \lesssim 10}$). Then the second threads are added to each core of the first processor (${10 < M\le 20}$). The next stage begins with the use of the second processor (${20 < M \le 40}$). This transition is especially noticeable for a very large number of particles in models with ${M \le M_P}$ (See figure~\ref{fig:k80-M}).
Comparing the efficiencies of $E^{(P)}(20)$ and $E^{(P)}(40)$ yields an increase of 77 percent for ${N = 10^5}$ and 8 percent for ${N = 10^8}$ for computer K80.

Figures~\ref{fig:m113-M} --- \ref{fig:m301-M} show the result of the efficiency calculation for systems with extra threads, as a continuation of the analysis for Epsilon and K80. The transition of the number of threads through ${M = M_P}$ remains critically important. Comparison of $E^{(P)}(8)$ and $E^{(P)}(16)$ for computer M105S gives a gain about of 12 percent when using the maximum number of extra threads. The result for computer M301is shown in figure~\ref{fig:m301-M}, which demonstrates a slightly different version of resource distribution among threads. We have a characteristic kink in the dependencies near ${M = M_P/2 = 8}$. This means that extra threads begin to be used, although there are still free physical cores. At the same time, using the maximum possible number of extra threads for the M301 processor also improves the efficiency of parallel code about by 40 percent in models with ${N \sim 10^5 - 10^6}$.

\begin{figure}[!htb]
\centering
\includegraphics[width = 0.85\textwidth]{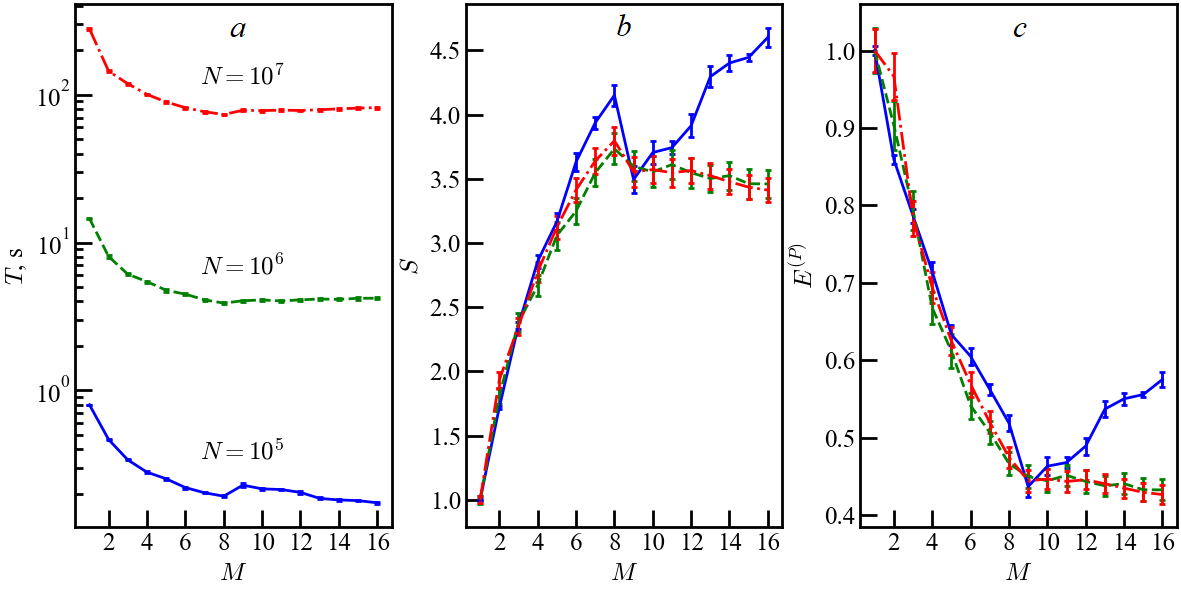}
\caption{
The same as in figure 1 for computer Ryzen. 
 }
\label{fig:ryzen-M}
\end{figure}

  %Компьютер Ryzen построен на базе AMD процессора (AMD processor). Результат вычисления $T(M)$, $S(M)$, $E^{(P)}(M)$ отличается от рассмотренных выше CPU (рисунок \ref{fig:ryzen-M}). Только модели с небольшим числом частиц воспроизводят аналогичные зависимости. Extra threads для моделей с ${N \ge 10^6}$ не дают прироста производительности, ухудшая результат при любом числе $M_E$. Является ли такой вывод общим для AMD processors или относится к конкретному CPU требует дополнительных изучений. 
Computer Ryzen is based on AMD processor. The result of calculating $T(M)$, $S(M)$, $E^{(P)}(M)$ differs from the CPUs considered above (figure~\ref{fig:ryzen-M}). Only models with a small number of particles reproduce similar dependencies. Extra threads for models with ${N \ge 10^6}$ do not provide a performance boost, worsening the result for any number of $M_E$. Whether this conclusion is general for AMD processors or applies to a specific CPU requires further study.

%На рисунках~\ref{fig-speedup-fram2fcpu_wo_boost}, \ref{fig-speedup-fram2fcpu_w_boost} показаны зависимости ускорения и эффективности от отношения эффективной частоты оперативной памяти $f_{RAM}$ к частоте процессора $f_{CPU}$ для четырех вычислительных потоков.

\begin{figure}[!htb]
\centering
\includegraphics[width = 0.93\textwidth]{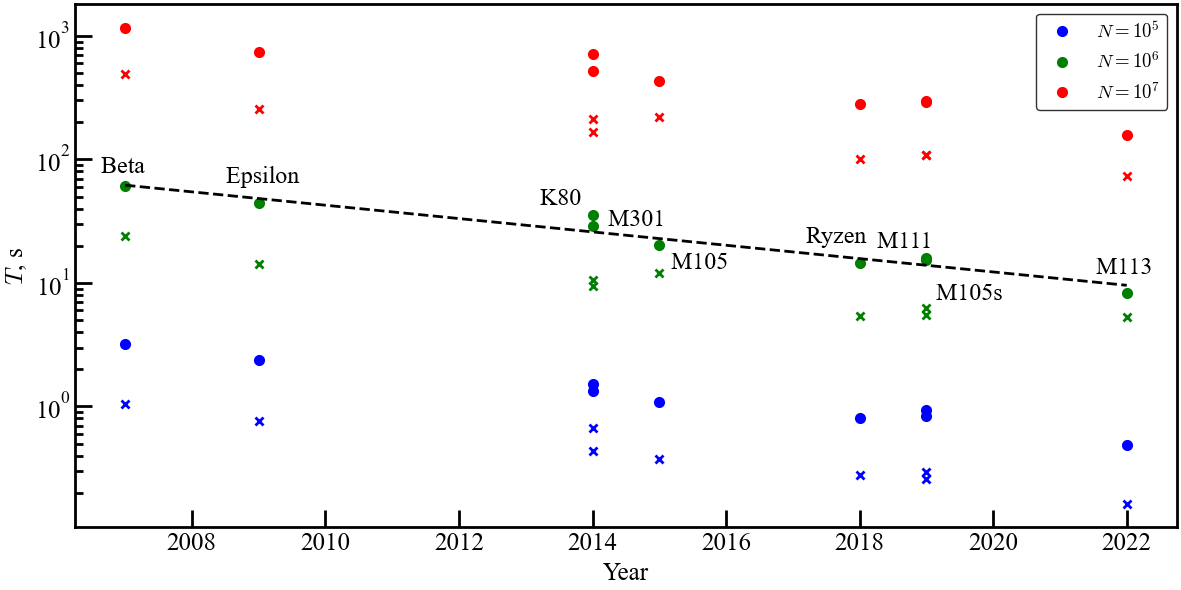}
\caption{
The execution time of one integration step on one core (${M = 1}$, circles) and four cores (${M = 4}$, crosses) from the processor release date for various computers from Table 1 in models with ${N = 10^5}$ (blue), ${N = 10^6}$ (green) and ${N = 10^7}$ (red). The black dashed line represents the approximation ${T \propto 2^{-t / \tau_y}}$ with ${\tau_y = 5.6}$ years.
 }
\label{fig:years-computers}
\end{figure}

Figure~\ref{fig:years-computers} shows the dependence of the program execution time on the processor release year from our sample. These results illustrate Moore's law. We have an approximately exponential decrease of the form ${T \propto 2^{-t / \tau_y}}$ with ${\tau_y = 5.6}$ years for the model with ${N = 10^6}$ and ${M = 1}$ (See the black dashed line in figure~\ref{fig:years-computers}). There is a noticeable spread of the time scale $\tau_y$ within 5--6 years for different models. The general trend of the data in figure~\ref{fig:years-computers} fits into the historical development of computing technology.
 
%\begin{figure}[!htb]
%\centering
%\includegraphics[width = 0.495\textwidth]{figs/fram2fcpu_wo_boost.png}
%\includegraphics[width = 0.495\textwidth]{figs/fram2fcpu_w_boost.png}
%\caption{Зависимость эффективности для четырех вычислительных потоков от отношения эффективной частоты оперативной памяти к частоте процессора для различных $N$: (a) для номинальной частоты процессора, (b) для частоты процессора в режиме <<Boost>>} 
%\label{fig-speedup-fram2fcpu}
%\end{figure}

\begin{figure}[!htb]
\centering
\includegraphics[width = 0.93\textwidth]{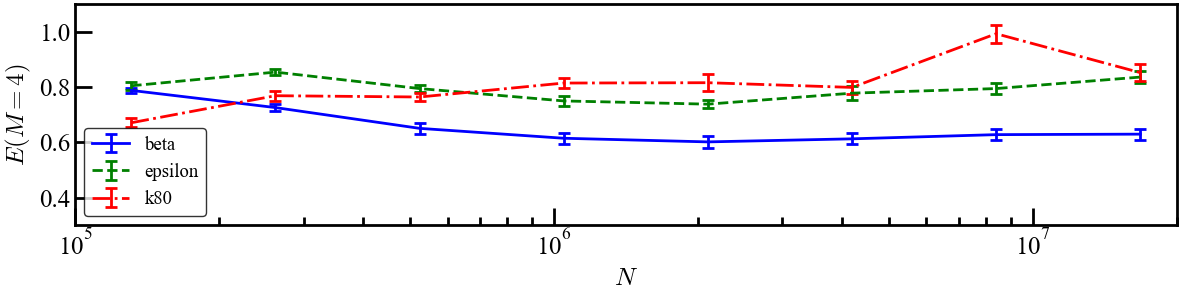}
\includegraphics[width = 0.93\textwidth]{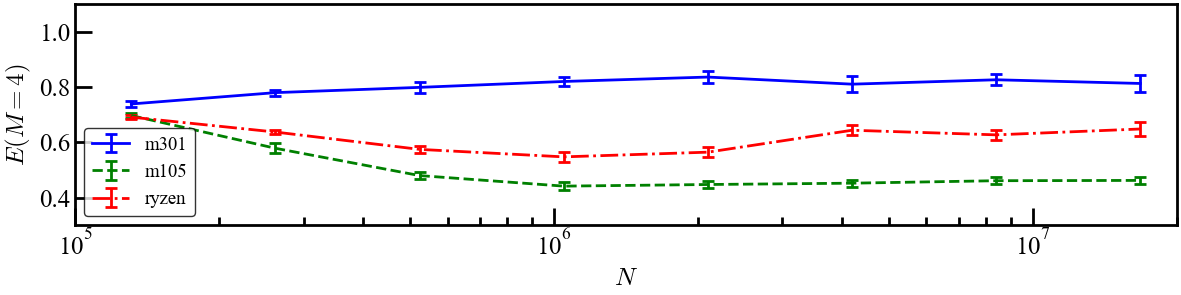}
\includegraphics[width = 0.93\textwidth]{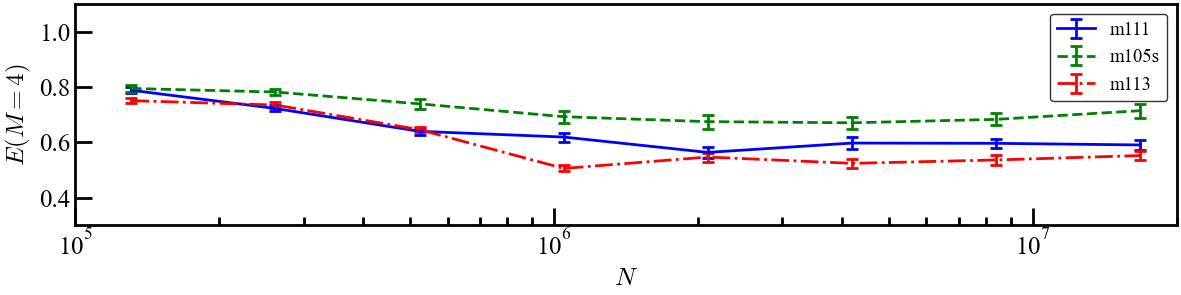}
\caption{
Dependence of efficiency on the number of particles for four computational threads for different processors. 
 }
\label{fig:nlogn}
\end{figure}

%\begin{figure}[!htb]
%\centering
%\includegraphics[width = \textwidth]{figs/nlogn_all.png}
%\caption{Зависимость эффективности от количества частиц для четырех вычислительных потоков для всех узлов {\color{red} Эта картинка только для нас, чтобы можно было увидеть сразу все}}
%\label{fig:nlogn-all}
%\end{figure}

The results discussed above are based on three main models with the number of particles ${N=10^5; 10^6; 10^7}$. Figure~\ref{fig:nlogn} shows in more detail the dependencies of the efficiency $E$ on the number of particles in the range ${2^{17} \le N \le 2^{24}}$. We restrict ourselves to the number of physical cores ${M = M_P = 4}$ and ${M_E = 0}$ in order to conveniently compare and cover all processors from Table 1.
Models with a small number of particles demonstrate a small spread of efficiency within ${0.67 - 0.81}$. Simulations with ${N = 2^{24} \simeq 1.7\cdot 10^7}$ give a spread of ${E = 0.45 - 0.85}$. An increase in $N$ can lead to both a decrease and an increase in efficiency, which is manifested in the non-monotonic behavior of the dependencies $E^{(P)}(N)$ in some models. Such a result was noted, for example, in simulations on GPU with CUDA for a similar model~\cite{Khrapov-Khoperskov-2024Frontiers}.

This effect can be caused by uneven distribution of data across RAM modules and dual-channel access to it. An important factor is also the ratio between the processor data processing speed ($P$) and RAM bandwidth ($R$), $P/R$. Let's consider two scenarios, in each of which all data are physically located in one RAM module with small $N$. The first case corresponds to the condition ${P/R > 1}$. Then the computation efficiency will grow with increasing $N$ until it reaches a certain value $N_{crit}$, after which it will begin to decrease. Further growth of $N$ requires placing data in two memory modules, which allows using the second access channel to it and leads to some increase in efficiency with a further plateau. The opposite inequality ${P/R < 1}$ determines the second option, when the condition ${N > N_{crit}}$ is satisfied immediately. The growth of $N$ is accompanied by some decrease in efficiency until the data is physically placed in two different memory modules. This allows the second channel to be used and the efficiency to increase slightly, reaching a plateau, as in the first case.

\section{Conclusion and discussion}

The analysis of the efficiency of parallel implementation of the code for calculating the gravitational force was performed for a system of $N$ particles using the Treecode method. We limited ourselves to considering the Open Multi-Processing (OpenMP) standard, since the use of the Message Passing Interface (MPI) for calculations on several processors significantly depends on the characteristics of the communication network.

Speedup and efficiency of parallel simulations depend on a large number of subtle factors that determine the architecture of multi-core processors. We focus on the role of additional logical threads (extra threads) that change the efficiency of executing parallel codes. Our results show that increasing the total number of threads due to extra threads ($M_E$) improves the efficiency of parallel Treecode calculations to an average 20 percent for different processors only in the case of ${M_E = M_P}$. If the number of extra threads is less than the number of physical cores, then the use of logical threads can significantly worsen the efficiency of parallelization.

A significant negative factor is the special reduction of the processor frequency in multi-core processors. The presence of a base frequency $\nu^{(1)}$ and a frequency in the Boost mode $\nu^{(2)}$ leads to calculations with a changing processor frequency. The transition from $\nu^{(1)}$ to $\nu^{(2)}$ can occur in several steps, since the processor frequency is adjusted to the current load. If the computer has several processors with shared memory, this can unbalance the efficiency of parallelization.

%Процессоры заточены (are designed) на снижение энергопотребления, которое пропорционально кубу от частоты процессора. Поэтому увеличение числа ядер со сниженной частотой может сохранить производительность при сильном уменьшении энергопотребления.

%Hyper-threading в случае Intel-процессоров и аналогичные технологии 

%Про кэш разных уровней

Note the emergence of processors with a new architecture, which have a more extensive set of cores. An example is processor Intel Core i9 14900K, which allows the use of 32 threads. The cores are divided into 8 performance cores (P-core) and 16 efficient cores (E-core). Performance cores support 8 extra threads, which gives a total number of threads ${M = 32}$. The question of the efficiency of such processors with a more complex architecture requires a separate analysis.

\begin{engbibliography}{99}
\bibitem{Athanassoula-1998N-Body}
%Athanassoula E. N-Body Simulations of Galaxies and Groups of Galaxies with the Marseille GRAPE Systems // Nonlinear Dynamics and Chaos in Astrophysics: A Festschrift in Honor of George Contopoulos. Edited by J. R. Buchler, S. T. Gottesman, and H. E. Kandrup. New York Academy of Sciences vol. 867, 1998., p.141-155. doi: 10.1111/j.1749-6632.1998.tb11255.x
\baut{Athanassoula}{E.}
\btit{N-Body Simulations of Galaxies and Groups of Galaxies with the Marseille GRAPE Systems}
\bj{Annals of the New York Academy of Sciences}
\bvol{867}
\bnum{1}
\bpp{141-155}
\bdoi{https://doi.org/10.1111/j.1749-6632.1998.tb11255.x}
\byr{1998}
\mkpapere

\bibitem{Bagla-2005Cosmological-N-Body}
%Bagla, J.S. Cosmological N-Body simulation: Techniques, Scope and Status // Current Science, 2005, Vol. 88, p. 1088-1100. doi: 10.48550/arXiv.astro-ph/0411043 
\baut{Bagla}{J. S.}
\btit{Cosmological N-Body simulation: Techniques, Scope and Status}
\bj{Current Science}
\bvol{88}
\bpp{1088-1100}
\bdoi{https://doi.org/10.48550/arXiv.astro-ph/0411043}
\byr{2005}
\mkpapere

\bibitem{Barnes-Hut-1986treecode}
%Barnes J. , Hut P.  (December 1986). A hierarchical O(N log N) force-calculation algorithm // Nature, 1986, 324 (4), 446–449
\baut{Barnes}{J.}
\baut{Hut}{P.}
\btit{A hierarchical O(N log N) force-calculation algorithm}
\bj{Nature}
\bvol{324}
\bnum{4}
\bpp{446-449}
\bdoi{https://doi.org/10.1038/324446a0}
\byr{1986}
\mkpapere

\bibitem{Bissekenov-etal-2024open-cluster}
%Cluster membership analysis with supervised learning and N-body simulations  / Bissekenov, A.; Kalambay, M.; Abdikamalov, E.; Pang, X.; Berczik, P.; Shukirgaliyev, B. // Astronomy \& Astrophysics, Volume 689, id.A282, 10 pp. doi: 10.1051/0004-6361/202449791
\baut{Bissekenov}{A.}
\baut{Kalambay}{M.}
\baut{Abdikamalov}{E.}
\baut{Pang}{X.}
\baut{Berczik}{P.}
\baut{Shukirgaliyev}{B.}
\btit{Cluster membership analysis with supervised learning and N-body simulations}
\bj{Astronomy \& Astrophysics}
\bvol{689}
\baid{A282}
\bdoi{https://doi.org/10.1051/0004-6361/202449791}
\byr{2024}
\mkpapere

\bibitem{Butenko-etal-2022MPCS}
\baut{Butenko}{M. A.}
\baut{Belikova}{I. V.}
\baut{Kuzmin}{N. M.}
\baut{Khokhlova}{S. S.}
\baut{Ivanchenko}{G. S.}
\baut{Ten}{A. V.}
\baut{Kudina}{I. G.}
\btit{Numerical simulation of the galaxies outer spiral structure: the influence of the dark halo non-axisymmetry on the gaseous disk shape}
\bj{Mathematical Physics and Computer Simulation}
\byr{2022}
\bvol{25 (3)}
\bpp{73-83}
\bdoi{https://doi.org/10.15688/mpcm.jvolsu.2022.3.5}
\mkpapere

\bibitem{Ciambur-etal-2021N-body-bar} 
%Double X/Peanut structures in barred galaxies - insights from an N-body simulation/ Ciambur, B.C.; Fragkoudi, F.; Khoperskov, S.; Di Matteo, P.; Combes, F.  // Monthly Notices of the Royal Astronomical Society, 2021, Volume 503, Issue 2, pp.2203-2214. doi: 10.1093/mnras/staa3814 
\baut{Ciambur}{B.C.}
\baut{Fragkoudi}{F.}
\baut{Khoperskov}{S.}
\baut{Di Matteo}{P.}
\baut{Combes}{F.}
\btit{Double X/Peanut structures in barred galaxies – insights from an N-body simulation}
\bj{Monthly Notices of the Royal Astronomical Society}
\bvol{503}
\bnum{2}
\bpp{2203-2214}
\bdoi{https://doi.org/10.1093/mnras/staa3814}
\byr{2020}
\mkpapere

\bibitem{Dubinski-etal-2004treecode}
\btit{GOTPM: a parallel hybrid particle-mesh treecode}
\baut{Dubinski}{J.}
\baut{Kim}{J.}
\baut{Park}{C.}
\baut{Humble}{R.}
\bj{New Astronomy}
\byr{2004}
\bvol{9}
\bpp{111-126}
\bdoi{https://doi.org/10.1016/j.newast.2003.08.002}
\mkpapere

\bibitem{Fattahi-etal-2016Apostle-project}
\baut{Fattahi}{A.}
\baut{Navarro}{J. F.}
\baut{Sawala}{T.}
\baut{Frenk}{C. S.}
\baut{Oman}{K. A.}
\baut{Crain}{R. A.}
\baut{Furlong}{M.}
\baut{Schaller}{M.}
\baut{Schaye}{J.}
\baut{Theuns}{T.}
\baut{Jenkins}{A.}
\btit{The Apostle project: Local Group kinematic mass constraints and simulation candidate selection}
\bj{Monthly Notices of the Royal Astronomical Society}
\byr{2016}
\bvol{457}
\bpp{844-856}
\bdoi{https://doi.org/10.1093/mnras/stv2970}
\mkpapere

\bibitem{Fridman-Khoperskov-2012book}
%Fridman A.M., Khoperskov A.V. Physics of Galactic Disks // Cambridge International Science Publishing Ltd. 2012, 754 p.
\baut{Fridman}{A. M.}
\baut{Khoperskov}{A. V.}
\btit{Physics of Galactic Disks}
\bpub{Cambridge International Science Publishing Ltd}
\byr{2012}
\bpp{754}
\mkbooke

\bibitem{Grand-etal-2016AURIGA-project}
\baut{Grand}{R. J. J.}
\baut{Springel}{V.}
\baut{Gomez}{F. A.}
\baut{Marinacci}{F.}
\baut{Pakmor}{R.}
\baut{Campbell}{D. J. R.}
\baut{Jenkins}{A.} 
\btit{Vertical disc heating in Milky Way-sized galaxies in a cosmological context}
\byr{2016}
\bj{Monthly Notices of the Royal Astronomical Society}
\bvol{459}
\bpp{199-219}
\mkpapere

\bibitem{Hernandez-Aguayo-etal-2023MillenniumTNG} 
\baut{Hernandez-Aguayo}{C.}
\baut{Springel}{V.}
\baut{Pakmor}{R.}
\baut{Barrera}{M.}
\baut{Ferlito}{F.}
\baut{White}{S. D. M.}
\baut{Hernquis}{L.}
\baut{Hadzhiyska}{B.}
\baut{Delgado}{A. M.}
\baut{Kannan}{R.}
\baut{Bose}{S.}
\baut{Frenk}{Carlos}
\btit{The MillenniumTNG Project: high-precision predictions for matter clustering and halo statistics}
\bj{Monthly Notices of the Royal Astronomical Society}
\byr{2023}
\bvol{524}
\bpp{2556-2578}
\bdoi{https://doi.org/10.1093/mnras/stad1657 }
\mkpapere

\bibitem{Ishchenko-etal-2024globular-clusters}
%Dynamical evolution of Milky Way globular clusters on the cosmological timescale: I. Mass loss and interaction with the nuclear star cluster / Ishchenko, Maryna; Berczik, Peter; Panamarev, Taras; Kuvatova, Dana; Kalambay, Mukhagali; Gluchshenko, Anton; Veles, Oleksandr; Sobolenko, Margaryta; Sobodar, Olexander; Omarov, Chingis   // Astronomy \& Astrophysics, 2024, Volume 689, id.A178, 17 pp. doi: 10.1051/0004-6361/202450399
\baut{Ishchenko}{M.}
\baut{Berczik}{P.}
\baut{Panamarev}{T.}
\baut{Kuvatova}{D.}
\baut{Kalambay}{M.}
\baut{Gluchshenko}{A.}
\baut{Veles}{O.}
\baut{Sobolenko}{M.}
\baut{Sobodar}{O.}
\baut{Omarov}{C.}
\btit{Dynamical evolution of Milky Way globular clusters on the cosmological timescale - I. Mass loss and interaction with the nuclear star cluster}
\bj{Astronomy \& Astrophysics}
\bvol{689}
\baid{A178}
\bdoi{https://doi.org/10.1051/0004-6361/202450399}
\byr{2024}
\mkpapere

\bibitem{Ishchenko-etal-2024Berczik}
%Ishchenko, M.; Berczik, P.; Sobolenko, M. Milky Way globular clusters on cosmological timescales. IV. Guests in the outer Solar System // Astronomy \& Astrophysics, 2024, Volume 683, id.A146, 11 pp. doi: 10.1051/0004-6361/202347990 
\btit{Milky Way globular clusters on cosmological timescales. IV. Guests in the outer Solar System}
\baut{Ishchenko}{M.}
\baut{Berczik}{P.}
\baut{Sobolenko}{M.}
\bj{Astronomy \& Astrophysics}
\bvol{683}
\baid{A146}
\bdoi{https://doi.org/10.1051/0004-6361/202347990}
\byr{2024}
\mkpapere

\bibitem{Just-etal-2023Polyachenko}
%Global survey of star clusters in the Milky Way. VII. Tidal parameters and mass function / Just, A.; Piskunov, A.E.; Klos, J.H.; Kovaleva, D.A.; Polyachenko, E.V. // Astronomy \& Astrophysics, 2023, Volume 672, id.A187, 18 pp. doi: 10.1051/0004-6361/202244723
\btit{Global survey of star clusters in the Milky Way - VII. Tidal parameters and mass function}
\baut{Just}{A.}
\baut{Piskunov}{A. E.}
\baut{Klos}{J. H.}
\baut{Kovaleva}{D. A.}
\baut{Polyachenko}{E. V.}
\bj{Astronomy \& Astrophysics}
\bvol{672}
\baid{A187}
\bdoi{https://doi.org/10.1051/0004-6361/202244723}
\byr{2023}
\mkpapere

\bibitem{Khoperskov-etal-2024Galaxies-cE}
%Khoperskov A.V., Khrapov S.S., Sirotin D.S. Formation of transitional cE/UCD galaxies through massive disc to dwarf galaxy mergers // Galaxies, 2024, v.12(1), 1, 1-29. https://doi.org/10.3390/galaxies12010001
\btit{Formation of transitional cE/UCD galaxies through massive disc to dwarf galaxy mergers}
\baut{Khoperskov}{A. V.}
\baut{Khrapov}{S. S.}
\baut{Sirotin}{D. S.}
\bj{Galaxies}
\bvol{12}
\bnum{1}
\baid{1}
\bdoi{https://doi.org/10.3390/galaxies12010001}
\byr{2024}
\mkpapere

\bibitem{Khrapov-Khoperskov-2024Frontiers} 
%Khrapov S., Khoperskov A. Study of the Effectiveness of Parallel Algorithms for Modeling the Dynamics of Collisionless Galactic Systems on GPUs // Supercomputing Frontiers and Innovations, 2024, Vol. 11, No. 3, 27-44   DOI: 10.14529/jsfi240302  
\btit{Study of the Effectiveness of Parallel Algorithms for Modeling the Dynamics of Collisionless Galactic Systems on GPUs}
\baut{Khrapov}{S.}
\baut{Khoperskov}{A.}
\bj{Supercomputing Frontiers and Innovations}
\bvol{11}
\bnum{3}
\bpp{27-44}
\bdoi{https://doi.org/10.14529/jsfi240302}
\byr{2024}
\mkpapere

\bibitem{Khrapov-etal-2023dwarf}
%Formation of spiral dwarf galaxies: observational data and results of numerical simulation / Khrapov S.S., Khoperskov A.V., Zaitseva N.A., Zasov A.V., Titov A.V., , St. Petersburg State Polytechnical University Journal. Physics and Mathematics. 16 (1.2) (2023) 395-402. DOI: https://doi.org/10.18721/JPM.161.260
\btit{Formation of spiral dwarf galaxies: observational data and results of numerical simulation}
\baut{Khrapov}{S. S.}
\baut{Khoperskov}{A. V.}
\baut{Zaitseva}{N. A.}
\baut{Zasov}{A. V.}
\baut{Titov}{A. V.}
\bj{St. Petersburg State Polytechnical University Journal. Physics and Mathematics}
\bpp{395-402}
\bvol{16}
\bnum{1.2}
\bdoi{https://doi.org/10.18721/JPM.161.260}
\byr{2023}
\mkpapere

\bibitem{Khrapov-etal-2018N-body} 
%Khrapov S.S., Khoperskov S.A., Khoperskov A.V. New features of parallel implementation of N-body problems on GPU // Bulletin of the South Ural State University, Series: Mathematical Modelling, Programming and Computer Software, 2018, v.11, No.1, p.124-136. DOI: 10.14529/mmp180111\\
\btit{New features of parallel implementation of N-body problems on GPU}
\baut{Khrapov}{S. S.}
\baut{Khoperskov}{S. A.}
\baut{Khoperskov}{A. V.}
\bj{Bulletin of the South Ural State University, Series: Mathematical Modelling, Programming and Computer Software}
\bvol{11}
\bnum{1}
\bpp{124-136}
\bdoi{https://doi.org/10.14529/mmp180111}
\byr{2018}
\mkproce

\bibitem{Kulikov-2016interaction-galaxies}
%Kulikov, I. ; Chernykh, I. ; Protasov, V. Mathematical modeling of formation, evolution and interaction of galaxies in cosmological context // Journal of Physics: Conference Series, 2016, Volume 722, Issue 1, article id. 012023
\btit{Mathematical modeling of formation, evolution and interaction of galaxies in cosmological context}
\baut{Kulikov}{I.}
\baut{Chernykh}{I.}
\baut{Protasov}{V.}
\bj{Journal of Physics: Conference Series}
\bvol{722}
\bnum{1}
\baid{012023}
\bdoi{https://dx.doi.org/10.1088/1742-6596/722/1/012023}
\byr{2016}
\mkproce

\bibitem{Kyziropoulos-etal-2015Parallel-N-Body}
%Kyziropoulos P. E. , C. K. Filelis-Papadopoulos, G. A. Gravvanis Parallel N-Body Simulation Based on the PM and P3M Methods Using Multigrid Schemes in conjunction with Generic Approximate Sparse Inverses // Mathematical Problems in Engineering, 2015, V. 2015, id.450980. https://doi.org/10.1155/2015/450980
\btit{Parallel N-Body Simulation Based on the PM and P3M Methods Using Multigrid Schemes in conjunction with Generic Approximate Sparse Inverses}
\baut{Kyziropoulos}{P. E.}
\baut{Filelis-Papadopoulos}{C. K.}
\baut{Gravvanis}{G. A.}
\bj{Mathematical Problems in Engineering}
\bvol{2015}
\baid{450980}
\bdoi{https://doi.org/10.1155/2015/450980}
\byr{2015}
\mkpapere

\bibitem{Huillier-etal-2014N-body}
%Huillier Benjamin L’Huillier, Changbom Park, Juhan Kim Effects of the initial conditions on cosmological N-body simulations // New Astronomy, 2014, Volume 30, Pages 79-88.
%https://doi.org/10.1016/j.newast.2014.01.007
\btit{GalaxyFlow: upsampling hydrodynamical simulations for realistic mock stellar catalogues}
\baut{L’Huillier}{H. B.}
\baut{Park}{C.}
\baut{Kim}{J.}
\bj{New Astronomy}
\bvol{30}
\bpp{79-88}
\bdoi{https://doi.org/10.1016/j.newast.2014.01.007}
\byr{2014}
\mkpapere

\bibitem{Lim-etal-2024GalaxyFlow}
%GalaxyFlow: upsampling hydrodynamical simulations for realistic mock stellar catalogues/ Lim, S.H.; Raman, K.A.; Buckley, M.R.; Shih, D.  // Monthly Notices of the Royal Astronomical Society, 2024, Volume 533, Issue 1, pp.143-164
\btit{GalaxyFlow: upsampling hydrodynamical simulations for realistic mock stellar catalogues}
\baut{Lim}{S. H.}
\baut{Raman}{K. A.}
\baut{Buckley}{M. R.}
\baut{Shih}{D.}
\bj{Monthly Notices of the Royal Astronomical Society}
\bvol{533}
\bnum{1}
\bpp{143-164}
\bdoi{https://doi.org/10.1093/mnras/stae1672}
\byr{2024}
\mkpapere

%\bibitem{Liu-Bhatt-2000parallel-N-body}
%Liu P., Bhatt S.N. Experiences with parallel N-body simulation, in IEEE Transactions on Parallel and Distributed Systems, vol. 11, no. 12, pp. 1306-1323, Dec. 2000, doi: 10.1109/71.895795 
%\btit{Experiences with parallel N-body simulation, in IEEE Transactions on Parallel and Distributed Systems}
%\baut{Liu}{P.}
%\baut{Bhatt}{S. N.}
%\bj{IEEE Transactions on Parallel and Distributed Systems}
%\bvol{11}
%\bnum{12}
%\bpp{1306-1323}
%\bdoi{https://doi.org/10.1109/71.895795}
%\byr{2000}
%\mkpapere

\bibitem{Nipoti-2021N-body} 
%Effective N-body models of composite collisionless stellar systems / Nipoti, C.; Cherchi, G.; Iorio, G.; Calura, F.  // Monthly Notices of the Royal Astronomical Society, 2021, Volume 503, Issue 3, pp.4221-4230. doi: 10.1093/mnras/stab763
\baut{Nipoti}{C.}
\baut{Cherchi}{G.}
\baut{Iorio}{G.}
\baut{Calura}{F.}
\btit{Effective N-body models of composite collisionless stellar systems}
\bj{Monthly Notices of the Royal Astronomical Society}
\bvol{503}
\bnum{3}
\bpp{4221-4230}
\bdoi{https://doi.org/10.1093/mnras/stab763}
\byr{2021}
\mkpapere

\bibitem{Pejch-etal-2023}
\baut{Pejch}{M. A.}
\baut{Morozov}{A. G.}
\baut{Khoperskov}{A. V.}
\btit{Modeling a double-hump gas rotation curves in the axisymmetric gravitational field of galaxies}
\bj{Mathematical Physics and Computer Simulation}
\byr{2023}
\bvol{3}
\bpp{91-104}
\mkpapere

\bibitem{Pillepich-etal-2018Illustris-TNG}
\baut{Pillepich}{A.}
\baut{Springel}{V.}
\baut{Nelson}{D.}
\baut{Genel}{S.}
\baut{Naiman}{J.}
\baut{Pakmor}{R.}
\baut{Hernquist}{L.}
\baut{Torrey}{P.}
\baut{Vogelsberger}{M.}
\baut{Weinberger}{R.}
\baut{Marinacci}{F.}
\btit{Simulating galaxy formation with the IllustrisTNG model}
\bj{Monthly Notices of the Royal Astronomical Society}
\byr{2018}
\bvol{473}
\bpp{4077-4106}
\bdoi{https://doi.org/10.1093/mnras/stx2656}
\mkpapere

\bibitem{Potter-etal-2017PKDGRAV3}
\baut{Potter}{D.}
\baut{Stadel}{J.}
\baut{Teyssier}{R.} 
\btit{PKDGRAV3: beyond trillion particle cosmological simulations for the next era of galaxy surveys}
\bj{Computational Astrophysics and Cosmology}
\byr{2017}
\bvol{4, id. 2}
\bdoi{https://doi.org/10.1186/s40668-017-0021-1}
\mkpapere

\bibitem{Ruan-etal-2022N-body-modified-gravity}
%Fast full N-body simulations of generic modified gravity: conformal coupling models / Cheng-Zong Ruan, Cesar Hernandez-Aguayo, Baojiu Li, Christian Arnold, Carlton M. Baugh, Anatoly Klypin, Francisco Prada //  Journal of Cosmology and Astroparticle Physics, 2022, V.2022, id.018 DOI 10.1088/1475-7516/2022/05/018
\baut{Ruan}{Cheng-Zong}
\baut{Hernandez-Aguayo}{C.}
\baut{Li}{B.}
\baut{Christian}{A.}
\baut{Carlton}{M. B.}
\baut{Klypin}{A.}
\baut{Prada}{F.}
\btit{Fast full N-body simulations of generic modified gravity: conformal coupling models}
\bj{Journal of Cosmology and Astroparticle Physics}
\bvol{2022}
\bnum{5}
\baid{018}
\bdoi{https://dx.doi.org/10.1088/1475-7516/2022/05/018}
\byr{2022}
\mkpapere

\bibitem{Schaye-etal-2015EAGLE-project}
\baut{Schaye}{J.}
\baut{Crain}{R. A.}
\baut{Bower}{R. G.}
\baut{Furlong}{F.}
\baut{Schaller}{M.}
\baut{Theuns}{T.}
\baut{Vecchia}{C. D.}
\baut{Frenk}{C. S.}
\baut{McCarthy}{I. G.}
\baut{Helly}{J. C.}
\baut{Jenkins}{A.}
\baut{Rosas-Guevara}{Y. M.}
\baut{White}{S. D. M.}
\baut{Baes}{M.}
\baut{Booth}{C. M.}
\baut{Camps}{P.}
\baut{Navarro}{J. F.}
\baut{Qu}{Y.}
\baut{Rahmati}{A.}
\baut{Sawala}{T.}
\baut{Thomas}{P.A.}
\baut{Trayford}{J.}
\btit{The EAGLE project: simulating the evolution and assembly of galaxies and their environments}
\bj{Monthly Notices of the Royal Astronomical Society}
\bvol{446}
\bpp{521-554}
\bdoi{https://doi.org/10.1093/mnras/stu2058}
\mkpapere

\bibitem{Schaye-etal-2023cosmolog}
\baut{Schaye}{J.}
\baut{Kugel}{R.}
\baut{Schaller}{M.}
\baut{Helly}{J. C.}
\baut{Braspenning}{J.}
\btit{The FLAMINGO project: cosmological hydrodynamical simulations for large-scale structure and galaxy cluster surveys}
\bj{Monthly Notices of the Royal Astronomical Society}
\byr{2023}
\bvol{526}
\bpp{4978-5020}
\bdoi{https://doi.org/10.1093/mnras/stad2419}
\mkpapere

\bibitem{Smirnov-etal-2017slow-bars}
\btit{Simulations of slow bars in anisotropic disk systems}
\baut{Smirnov}{A. A.}
\baut{Sotnikova}{N. Y.}
\baut{Koshkin}{A. A.}
\bj{Astronomy Letters}
\byr{2017}
\bvol{43}
\bpp{61-74}
\mkpapere

\bibitem{Springel-etal-2001GADGET}
\baut{Springel}{V.}
\baut{Yoshida}{N.}
\baut{White}{S. D. M.}
\btit{GADGET: a code for collisionless and gasdynamical cosmological simulations}
\bj{New Astronomy}
\byr{2001}
\bvol{6}
\bpp{79-117}
\bdoi{https://doi.org/10.1016/S1384-1076(01)00042-2}
\mkpapere

\bibitem{Tikhonenko-etal-2021Sotnikova}
\baut{Tikhonenko}{I. S.}
\baut{Smirnov}{A. A.}
\baut{Sotnikova}{N. Ya.}
\btit{First direct identification of the barlens vertical structure in galaxy models}
\bj{Astronomy \& Astrophysics}
\byr{2021}
\bvol{648}
\bpp{5}
\bdoi{https://doi.org/10.1051/0004-6361/202140703}
\mkpapere

\bibitem{Titov-Khoperskov-2022Peter} 
\btit{Numerical Modeling of the Collisions of Spheroidal Galaxies: Mass Loss Efficiency by Baryon Components}
\baut{Titov}{A. V.}
\baut{Khoperskov}{A. V.}
\bj{Vestnik St. Petersburg University, Mathematics}
\bvol{55}
\bnum{1}
\bpp{124-134}
\byr{2022}
\bdoi{https://doi.org/10.1134/S1063454122010149}
\mkpapere
 
\bibitem{Vogelsberger-etal-2020Cosmological-simulations}
%Cosmological simulations of galaxy formation/ Vogelsberger M., Marinacci F., Torrey P., Puchwein E.  // Nature Reviews Physics, 2020, Volume 2, Issue 1, p.42-66. doi: 10.1038/s42254-019-0127-2 
\baut{Vogelsberger}{M.}
\baut{Marinacci}{F.}
\baut{Torrey}{P.}
\baut{Puchwein}{E.}
\btit{Cosmological simulations of galaxy formation}
\bj{Nature Reviews Physics}
\bvol{2}
\bpp{42-66}
\bdoi{https://doi.org/10.1038/s42254-019-0127-2}
\byr{2020}
\mkpapere

\bibitem{Walther-2003P3M}
\baut{Walther}{J. H.}
\btit{An influence matrix particle–particle particle-mesh algorithm with exact particle–particle correction}
\bj{Journal of Computational Physics}
\bvol{184}
\byr{2003}
\bpp{670-678}
\mkpapere

\bibitem{Yokota-Barba-2011Fast-Multipole-Method}
\baut{Yokota}{R.}
\baut{Barba}{L. A.}
\btit{Treecode and Fast Multipole Method for N-Body Simulation with CUDA}
\bj{GPU Computing Gems Emerald Edition. Applications of GPU Computing Series}
\byr{2011}
\bpp{113-132}
\bdoi{https://doi.org/10.1016/B978-0-12-384988-5.00009-7}
\mkpapere

\end{engbibliography}

\selectlanguage{russian}

%\begin{summary}
%Моделирование бесстолкновительных галактических систем основывается на модели N-тел, которая требует больших вычислительных ресурсов из-за дальнодействующего характера гравитационных сил. Наиболее распространенным методом вычисления гравитации является приближенный алгоритм TreeCode, обеспечивающий более быстрое вычисление силы по сравнению с прямым суммированием вкладов от всех частиц. Проведен анализ вычислительной эффективности для моделей с числом частиц в пределах $10^{8}$. Мы рассмотрели несколько процессоров с различной архитектурой с целью определения производительности параллельных симуляций на основе стандарта OpenMP. Анализ использования дополнительных потоков (extra threads) дополнительно к физическим ядрам показывает рост производительности симуляций только при загрузке всех логических потоков. Это дает прирост параллельной эффективности (efficiency of parallel computing) в среднем на 20 процентов. 
%\end{summary}

\end{document}